\documentclass[a4paper,aps,prb,twocolumn,floatfix]{revtex4-1}
\usepackage{epsfig}
\usepackage{amsmath}
\usepackage{graphics}
\usepackage{graphicx}

\begin{document}

\title{Effects of vertex corrections on diagrammatic approximations applied to  the study of transport through a quantum dot}

\author{Leandro Tosi$\sp{1}$}
\author{Pablo Roura-Bas$\sp{2}$}
\author{Ana Mar\'{\i}a Llois$\sp{2}$}
\author{Luis O. Manuel$\sp{3}$}

\affiliation{$\sp{1}$Centro At\'{o}mico Bariloche and}
\affiliation{$\sp{2}$Centro At\'omico Constituyentes, Comisi\'on Nacional de Energ\'ia At\'omica, Argentina}
\affiliation{$\sp{3}$Instituto de F\'isica Rosario, Universidad Nacional de Rosario, Rosario, Argentina}

\email{roura@tandar.cnea.gov.ar}

\begin{abstract}

In the present work, we calculate the conductance through a single quantum dot weakly coupled to metallic contacts. 
We use the spin-1/2  Anderson model to describe the quantum dot, while considering a finite Coulomb repulsion. 
We solve the interacting system using the non-crossing-approximation (NCA) and the one-crossing 
approximation (OCA). We obtain the linear response conductance as a function of temperature and energy position of 
the localized level.  From the comparison of  both approximations  we extract  the role of the  vertex corrections, 
which are introduced in the OCA calculations and neglected in the NCA scheme. As a function of the energy position, 
we observe that the diagrams omitted within NCA are really important for appropriately describing transport phenomena 
in Kondo systems as well as in the mixed valence regime. On the other hand, as a function of temperature, the corrections 
introduced by OCA partly recover the universal scaling properties known from numerical approaches such as the Numerical 
Renormalization Group(NRG).  

\end{abstract}

\maketitle

\section{Introduction} 

Large-$N$ expansions, with $N$ representing the angular momentum degeneracy, are commonly used for solving the 
Anderson Impurity model (AIM) and to study the Kondo physics.
In particular, the so-called non-crossing approximation (NCA) in its infinite Coulomb repulsion limit captures the 
formation of the Kondo resonance at finite temperatures.\cite{Bickers}  When compared with the numerical renormalization 
group (NRG), the NCA scheme also provides the correct Kondo temperature ($T_K$).\cite{costi} This successful match results  
from the fact that NCA collects, in a self-consistent way, all diagrams up to order $1/N$. 
However, it is well known that the Fermi liquid properties are not properly described. To deal with this 
deficiency, infinite resummations of spin flip terms are required.\cite{anders, kroha, kirchner} 

When a finite value of Coulomb repulsion $U$ is allowed, the NCA gives a severely 
underestimated value of $T_K$.\cite{haule_1} The leading crossing diagrams that restore the proper energy scale were 
introduced in the early work by Pruschke and Grewe within the framework of the enhanced-NCA (ENCA) also called One 
Crossing Approximation (OCA).\cite{Pruschke} Unlike NRG and Quantum Monte Carlo, which are actually more robust 
techniques, the OCA method can be extended to multi orbital systems without much numerical effort. It also works 
on the real axis and can go to temperatures far below the Kondo one. These features make OCA a powerful impurity 
solver when combined with electronic structure calculations within the context of the Dynamical Mean Field Theory 
(DMFT).\cite{haule_2} 

Due to the interesting advantages mentioned above, it is important to check the role of vertex corrections on 
diagrammatic techniques in the calculations of different properties. Recently, the influence of vertex functions 
incorporated by OCA has already been studied for general lattice problems,\cite{grewe_2} multi-orbital Anderson 
models,\cite{grewe_1} and dynamic susceptibilities of the AIM, \cite{schmitt} among others. 

In this work, we study the effect of vertex corrections on transport properties, in particular transport through a 
quantum dot (QD) weakly coupled to metallic contacts. The NCA results for the conductance, in the linear-response 
regime, were previously analyzed by Gerace \textit{et al.}.\cite{gerace} 
The aim of our contribution is to compare the equilibrium conductance, as a function of temperature and gate voltage, 
as calculated by NCA and OCA methods. Furthermore, we analyze the scaling properties of the conductance by comparison 
with the empirical formula extracted from NRG calculations.   

We observe that the diagrams omitted within NCA are really important for appropriately describing transport phenomena. 
Furthermore, the vertex corrections introduced by OCA partially recover the universal scaling behavior of the 
conductance as a function of temperature.

The paper is organized as follows. In Sec. \ref{model} we introduce the model and the calculation methods. In 
Sec. \ref{results} we present and discuss numerical calculations. Finally, in Sec. \ref{conclusions} some conclusions 
are drawn.

\section{Model and method} \label{model}

To describe the system in which the QD is coupled to two leads we use the spin $-1/2$ Anderson Hamiltonian
\begin{eqnarray}\label{anderson}\begin{split}
H=& \sum_{k\nu\sigma}\varepsilon_{k\nu\sigma} c^{\dagger}_{k\nu\sigma} 
     c_{k\nu\sigma} + \sum_{\sigma}E_{d} n_{d\sigma} + U n_{d\uparrow}n_{d\downarrow}+ \\
& \sum_{k\nu\sigma}(V_{k\nu} d^{\dagger}_{\sigma} c_{k\nu\sigma}+ h.c.) ,\\
\end{split}\end{eqnarray}
where the operator $c^{\dagger}_{k\nu\sigma}$ represents a conduction electron with momentum $k$ and spin $\sigma$ 
in lead $\nu$, where $\nu=L,R$ labels the left and right leads; the operator $d^{\dagger}_{\sigma}$ stands for an 
electron in the dot and $n_{d\sigma}$ is the number operator for a given spin in the QD. The parameters $E_{d}$ and 
$U$ represent the energy of a single electron and the Coulomb repulsion in the dot, respectively. The coupling 
parameter and hybridization functions between leads and QD are given by $V_{k\nu}$ and 
$\Delta_{\nu}(\omega)\equiv\pi\sum_{\kappa\sigma} V_{\kappa\nu}^{2}\delta(\omega - \epsilon_{k\nu\sigma})$,
respectively.

\begin{figure}[h!]
\centering
\resizebox{6.cm}{!}{
\includegraphics[clip]{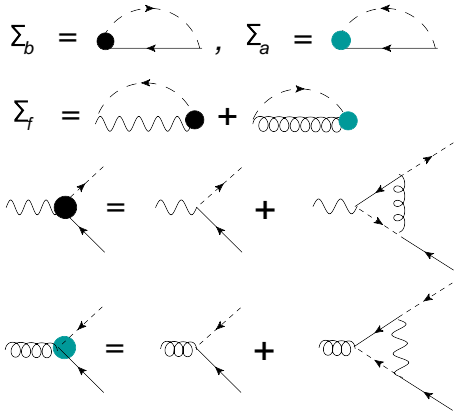}}
\caption{\small{Sketch of the self-consistency OCA scheme for self-energies and vertex corrections of the auxiliary
particles. 
The full, wiggly, dashed, and curly lines stand for pseudofermion, empty boson, conduction electron, and doubly
occupied 
boson propagators, respectively.}}
\label{1.eps}
\end{figure}

Our starting point is an auxiliary particle representation of the Hamiltonian given by 
eq. (\ref{anderson}).\cite{coleman} In this representation, the physical operator is given by the following combination 
of the auxiliary particles, ${d}^{\dagger}_{\sigma}={f}^\dagger_{\sigma}{b}+\sigma {a}^\dagger {f}_{-\sigma}$. 
Operators $b$, $f_\sigma$ and $a$ label vacuum, single, and doubly occupied states, respectively. Within the OCA scheme, 
the self-energies of the auxiliary particles and the vertex functions are obtained self-consistently. The set of OCA 
equations are sketched in Fig. \ref{1.eps}.\cite{Pruschke}
 
Once the spectral functions of the auxiliary particles are obtained, the physical spectral function, 
$\rho_d(\omega)=-1/\pi \mathcal{I} G_d(\omega)$, 
follows from direct convolution of the auxiliary ones and the vertex functions, where $G_d(\omega)$ represents the 
retarded Green function of the QD.

It should be pointed out that the NCA equations are recovered from OCA when the vertex corrections are neglected in 
the self-energies and physical Green's functions bubbles.

As already stated our aim in this work is to analyze the role of the crossing diagrams on equilibrium transport
properties. 
In particular, on the conductance through a QD within the linear-response regime which is given by\cite{win}  
\begin{eqnarray}\label{conductancia}
G = 4\frac{e^{2}}{h}\pi \sum_{\sigma} \int_{-\infty}^{\infty}~
d\omega \left[ -f'(\omega)\right]\Delta_{T}(\omega)\rho_d(\omega).  
\end{eqnarray}
In eq. (\ref{conductancia}) $\Delta_{T}(\omega)=\Delta_{L}(\omega)\Delta_{R}(\omega)/\Delta(\omega)$, where 
$\Delta(\omega)=\Delta_{L}(\omega)+\Delta_{R}(\omega)$ represents the total coupling between leads and QD, and
$f(\omega)=1/(1+e^{\beta\omega})$ is the Fermi function.

\section{Numerical Results} \label{results}

In this section, we present the results obtained for spectral density and conductance as a function of temperature 
and position of the QD level within, both, NCA and OCA schemes. When solving the self-consistency equations, we 
consider the case of
symmetric couplings, $\Delta_{L}=\Delta_{R}$, and the hybridization functions to be step functions
$\Delta_{\nu}(\omega)=\Delta_{\nu} \theta(D-\vert\omega\vert)$
with bandwidth $D$ several times larger than $\Delta_{\nu}$. From now on, we choose the total hybridization $\Delta=1$ 
as our unit of energy.

The numerical procedure used to solve the NCA and OCA equations, follows the method proposed by 
Kroha \textit{et al.}\cite{hettler} to describe the spectral densities of the auxiliary and physical particles. 
The procedure guarantees the resolution of the sets of integral equations for the auxiliary particles' self-energies 
to a high degree of accuracy down to very low temperatures. 

\begin{figure}[h!]
\centering
\resizebox{6.cm}{!}{
\includegraphics[clip]{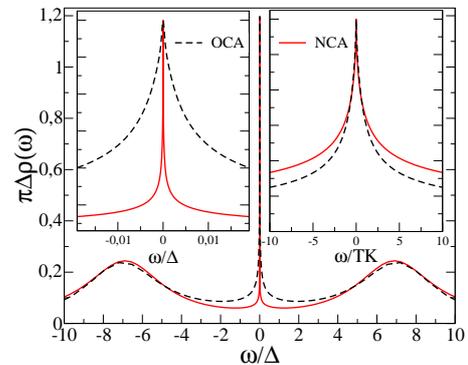}}
\caption{\small{Physical spectral function of the QD for the symmetric case, $E_d=-U/2=-6$ and $D=10$. 
The insets show a zoom of the region near the Fermi energy $\omega=0$.}}
\label{2.eps}
\end{figure}

As a first step, we obtain the physical spectral function for the symmetric Anderson model ($E_d=-U/2$) within NCA and 
OCA. Fig. \ref{2.eps} shows the results for $E_d=-6$ at the same temperature $T=0.01T_K$ in units of the 
corresponding $T_K$, being $T_K^{NCA}=0.00012$ and $T_K^{OCA}=0.007$. 
For the definition of Kondo temperature, $T_K$, we adopt the temperature for which $G(T_K)\equiv G_0/2$ with
$G_0=2\frac{e^{2}}{h}$.

As it was shown previously by Haule \textit{et al.},\cite{haule_1} from Fig. \ref{2.eps}, it is 
clear that the charge transfer peaks, located around $E_d$ and $ E_d+U$, are insensitive to vertex corrections. On the
other hand, 
the width of the Kondo resonance peak (left inset of Fig. \ref{2.eps}), which also represents the Kondo temperature,
increases 
orders of magnitude within OCA as compared to NCA. The larger value of $T_K$ provided by OCA can be traced back to
the additional spin-flip 
processes incorporated by the vertex corrections. In addition, when scaled by the corresponding $T_K$, the NCA
resonance is broader than the OCA one for energies beyond $T_K$ (right inset of Fig. \ref{2.eps}). 

\begin{figure}[h!]
\centering
\resizebox{6.cm}{!}{
\includegraphics[clip]{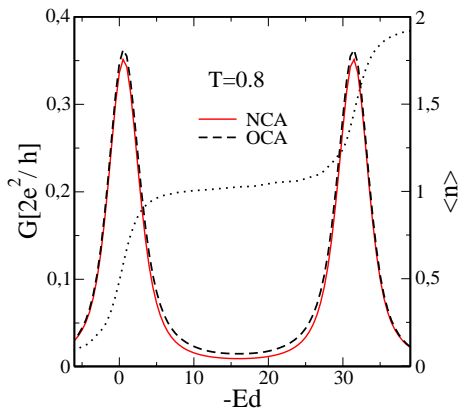}}
\caption{\small{Conductance in the linear-response regime and occupation (dotted curve) 
of QD at high temperature, $T=0.8$, as a function of the energy position $E_d$. 
Parameters: $U=32$, $D=43$. }}
\label{3.eps}
\end{figure}

We are interested here in the modifications induced on the conductance by the enhancement of the Kondo 
scale and the different shape of the Kondo resonance when scaled by the corresponding $T_K$.  
In Fig. \ref{3.eps} we present the conductance as a function of the energy level $E_d$, which is proportional to gate
voltage, 
in NCA and OCA schemes at a high temperature, $T=0.8$. It is also shown in Fig. \ref{3.eps} the NCA occupation of
the QD as a function of $E_d$. We observe that there is no difference between NCA and OCA occupations. This 
results from the fact that this magnitude is a static one, i.e. energy-integrated, and therefore weakly dependent 
on temperature. The value of the occupation number indicates the different regimes of the QD. 
In particular, $\langle n\rangle\sim 1$ stands for the Kondo regime. As it is seen, at this temperature value, 
there is no formation of the Kondo resonance within NCA nor within OCA.  
The conductance presents two separate peaks at the gate voltages for which the charge transfer peaks traverse the Fermi
level. 
Due to the absence of the Kondo peak at this temperature (see Fig. 1 of Ref.\cite{gerace}) there is no qualitative 
difference between NCA and OCA conductances, further reinforcing what we pointed in the spectral density of Fig. \ref{2.eps}. 

When the temperature decreases, due to the formation of the Kondo resonance at the Fermi level, the conductance valley
showed in Fig. \ref{3.eps} tends to raise as soon as the occupation in the QD is close to one. This means within the 
Kondo regime.

\begin{figure}[h!]
\centering
\resizebox{8.cm}{!}{
\includegraphics[clip]{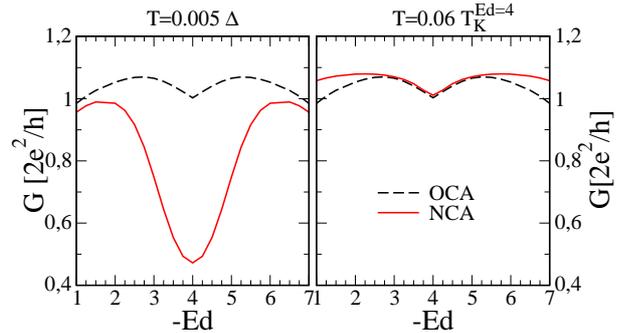}}
\caption{\small{Conductance through the QD at low temperatures as a function of the energy position $E_d$.
Parameters: $U=8$, $D=10$}.}
\label{4.eps}
\end{figure}

For low enough temperatures as compared to the Kondo one, the Kondo effect is fully developed and the Kondo peak 
of the spectral density reaches the limit imposed by Friedel's sum rule 
($\rho(0)\approx 1/\pi\Delta$). As a consequence, the conductance valley evolves into a plateau at $G_0$. 
This feature of the conductance is 
shown in the left panel of Fig. \ref{4.eps} for $T=0.005$ within OCA and is in agreement with the behavior
found by solving the spectral density of the QD using the slave bosons approach and NRG.\cite{armando_1}

However, as it is clear from Fig. \ref{2.eps}, NCA results differ from OCA ones at low temperatures.
Indeed, as we show in the left panel of Fig. \ref{4.eps}, due to the absence of vertex corrections the NCA
conductance still presents 
a valley or dip for the same temperature as compared to OCA. 
To obtain a similar plateau within NCA, using the same set of parameters, it would be necessary to go to temperatures 
several orders of magnitude smaller than within OCA. 
When compared at the same temperature in units of each $T_K$ for the symmetric case, right
panel of Fig. \ref{4.eps}, we observe a similar behavior of
both schemes near this $E_d$ value. In this case, we present the calculations at $T=0.06~T_K$, for which the
unitary limit is not exceeded within the Kondo regime (symmetric case). However, when some charge fluctuations are
allowed, we obtain a
larger violation of the unitary limit for the NCA conductance as compared to the OCA one. This represents an important
improvement of the OCA performance over the NCA one.

\begin{figure}[h!]
\centering
\resizebox{6.cm}{!}{
\includegraphics[clip]{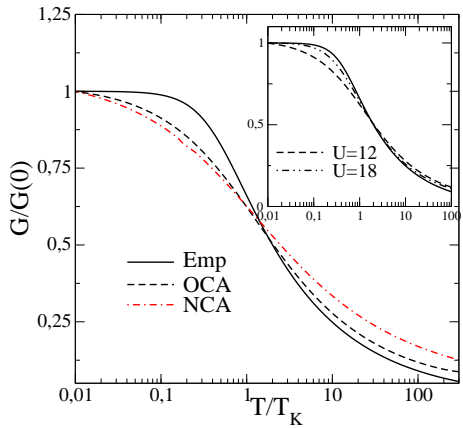}}
\caption{\small{Conductance as a function of $T/T_K$ for the symmetric Anderson model.
Parameters: $E_d = −6$, $U = 12$, $D = 10$.
The solid, dashed, and dot-dashed lines stand for NRG, OCA, and NCA techniques, respectively.
The inset shows the OCA conductance for the symmetric case, $U=12$, 
and far from this, $U=18$ (dot-dot-dashed line). }}
\label{5.eps}
\end{figure}

As we mentioned previously, when comparing the spectral function in units of $T_K$ (right inset of
Fig.
\ref{2.eps}) and
the conductance as a function of $E_d$ for the same temperature in units of $T_K$ (right panel of Fig. \ref{4.eps}),
NCA and OCA exhibit different scaling features. Thereafter, in what follows, we focus on the study of the universal
scaling properties of the conductance as a function of
temperature. 
NCA results for the universal behavior of the conductance as a function of $T/T_K$, were previously analyzed for
different values
of the energy level $E_d$.\cite{gerace} Also for different values of Coulomb repulsion $U$ close to the limit
$U\rightarrow\infty$\cite{roura}. In Fig. \ref{5.eps} we present the results for the conductance as a function
of $T/T_K$ for the symmetric case within the Kondo regime and at very low temperatures.  
In addition, we include the empirical formula for $G(T)$ extracted from calculations of the numerical 
renormalization group, NRG, for spin 1/2
\begin{equation}
G_{E}(T)=\frac{G(0)}{\left[ 1+(2^{1/s}-1)(T/T_{K})^{2}\right] ^{s}},
\label{ge}
\end{equation}
with $s=0.22$.\cite{G_E}
  
Due to the overestimation of Friedel's sum rule within both diagrammatic schemes (see Fig. \ref{2.eps}) in
Fig. \ref{5.eps} we normalize the conductance to $G(0)\equiv G(T_0)$, where $T_0\sim0.01T_K$ is the lowest temperature that
can be reached without obtaining unphysical results. Below this temperature, we observe that the conductance still
increases when a saturation is expected and therefore it can be considered as the limit of accuracy for
calculating the conductance.

We find that the NCA and OCA conductances follow different scaling behaviors in the whole range of temperatures. 
The OCA conductance is closer to the exact prediction of the NRG than NCA for both, low and high temperatures.  
In particular, an important improvement is obtained for temperatures larger than $T_K$. It must be noticed 
that vertex corrections not only change the Kondo scale but also the whole temperature dependence. 
However, it can be observed from Fig. \ref{5.eps} that the vertex functions introduced by OCA are not sufficient to 
recover the 
NRG prediction. As a last remark, it must be pointed out that while the NCA in its $U\rightarrow\infty$ limit
follows the
empirical law given by eq. (\ref{ge}),\cite{roura} we observe that the finite $U$ versions presented here fail
for the symmetric case. At this point, a set of self-consistent
vertex equations is still needed. 
The inset of Fig. \ref{5.eps} shows the conductance for the symmetric case with $U=12$ and far from this, $U=18$.
We notice that for larger values of $U$, the extra diagrams not included within OCA give a small contribution 
and as a consequence the conductance agrees very well with the NRG prediction.

\section{Conclusions} \label{conclusions}
In this work we obtain the conductance as a function of both, temperature and gate voltage, for a 
quantum dot modeled by the Anderson impurity Hamiltonian.
We calculate the spectral density of the quantum dot by using the finite Coulomb repulsion Non Crossing and
One Crossing Approximations. The comparison of both schemes let us conclude about the role of 
vertex corrections when calculating transport properties. 

At high temperatures, as compared with the hybridization strength, there is no qualitative difference between NCA 
and OCA conductances as a function of gate voltage as well as between the occupations of the dot.  
On the contrary, at low temperatures, the OCA conductance as a function 
of gate voltage displays a plateau, while the NCA one still shows a dip in the Kondo regime. 
This results from the underestimated Kondo scale within NCA. 

When studying the scaling properties, we find that the additional processes incorporated by the vertex 
functions modify the functional 
temperature dependence of the conductance in addition to the Kondo temperature. Thus, NCA and OCA show different
scaling 
behaviors in the whole range of temperatures.

Finally, when compared with the NRG results, the OCA conductance as a function of temperature is more reliable than 
the NCA one. 
For large values of the Coulomb repulsion, away from the symmetric case, the OCA and NRG conductances agree 
very well. However, close to the symmetric case, in 
order to recover the NRG prediction, it is still necessary to go beyond OCA corrections.

\section{ACKNOWLEDGMENTS}
Two of us (AML and LOM) are supported by CONICET. This work was done within the framework
of projects PIP 5254 and PIP 6016 of CONICET, and PICT 2006/483 and 33304 of
the ANPCyT and UBACyT X123.


\begin{thebibliography}{50}

\bibitem{Bickers} N.E. Bickers, Rev. Mod. Phys. \textbf{59}, 845 (1987).

\bibitem{costi} T. A. Costi, J. Kroha, and P. W\"{o}lfle, Phys. Rev. B \textbf{53}, 1850 (1996).

\bibitem{anders} F. B. Anders, J. Phys Condens. Matter \textbf{7}  2801 (1995).

\bibitem{kroha} J. Kroha, P. W\"olfle and T. A. Costi, Phys. Rev. Lett. \textbf{79}, 261 (1997).

\bibitem{kirchner} S. Kirchner, J. Kroha, and P. W\"olfle,  Phys. Rev. B   \textbf{70}, 165102 (2004).

\bibitem{haule_1}  K. Haule, S. Kirchner, J. Kroha, and P. W\"olfle,  Phys. Rev. B   \textbf{64}, 155111 (2001).

\bibitem{Pruschke} Th. Pruschke and N. Grewe, Z. Phys. B \textbf{74}, 439 (1989).

\bibitem{haule_2} K. Haule, C.H. Yee, and K. Kim, Phys. Rev. B \textbf{81}, 195107 (2010). 

\bibitem{grewe_2} N. Grewe, S. Schmitt, T. Jabben, and F. B. Anders, J. Phys.: Condens. Matter \textbf{20}, 365217 (2008).

\bibitem{grewe_1} N. Grewe, T. Jabben, and S. Schmitt,  Eur. Phys. J. B {\bf 68}, 23 (2009). 

\bibitem{schmitt} S. Schmitt, T. Jabben and N. Grewe, Phys. Rev. B \textbf{80}, 235130 (2009). 

\bibitem{gerace} D. Gerace, E. Pavarini, and L. C. Andreani, Phys. Rev. B \textbf{65}, 155331 (2002).

\bibitem{coleman}P. Coleman, Phys. Rev. B \textbf{29}, 3035 (1984).

\bibitem{win} N. S. Wingreen and Y. Meir, Phys. Rev. B \textbf{49}, 11040 (1994).

\bibitem{hettler} M.H. Hettler, J. Kroha, and S. Hershfield, Phys. Rev. B \textbf{58}, 5649 (1998).

\bibitem{armando_1} A. M. Lobos and A. A. Aligia, Phys. Rev. B \textbf{74}, 165417 (2006).

\bibitem{roura}  P. Roura-Bas, Phys. Rev. B \textbf{81}, 155327 (2010).

\bibitem{G_E} T. A. Costi, Phys. Rev. Lett. \textbf{85}, 1504 (2000).

\end{thebibliography}
\end{document}